\documentclass[11pt,a4paper]{article}

\usepackage{epsfig}
\usepackage{amssymb}
\usepackage{graphicx}
\usepackage{color}
\usepackage{subfigure}
\usepackage{mathtools}
\usepackage[hidelinks]{hyperref}

\makeatletter

\@addtoreset{equation}{section} \makeatother
\renewcommand{\vec}[1]{\underline{#1}}
\newcommand{\lra}[1]{\la{#1}\ra}

\def\a{\alpha}

\def\d{\delta}
\def\D{\Delta}

\def\ph{\phi}

\def\m{\mu}

\def\S{\Sigma}

\def\lt{\left}
\def\rt{\right}

\def\nn{\nonumber}

\def\p{\partial}

\def\la{\langle}
\def\ra{\rangle}

\def\nn{\nonumber}

\def\bea{\begin{eqnarray}}
\def\eea{\end{eqnarray}}

\begin{document}

\begin{titlepage}
\title{\vskip -60pt
\vskip 20pt Vacua, walls and junctions in $G_{N_F,N_C}$
}
\author{
Sunyoung Shin \footnote{e-mail:sihnsy@skku.edu}}
\date{}
\maketitle \vspace{-1.0cm}
\begin{center}
~~~
\it Institute of Basic Science, Sungkyunkwan University,\\
          Suwon 16419, Republic of Korea
\end{center}

\thispagestyle{empty}

\begin{abstract}
We discuss vacua, walls and three-pronged junctions of the mass-deformed nonlinear sigma models on the Grassmann manifold $G_{N_F,N_C}=\frac{SU(N_F)}{SU(N_C)\times SU(N_F-N_C)\times U(1)}$, which are non-Abelian gauge theories for $N_C\geq 2$. Polyhedra are proposed in \cite{Eto:2005cp} to describe Bogomol'nyi-Prasad-Sommerfield objects of the mass-deformed nonlinear sigma models on the complex projective space, which are Abelian gauge theories. We show that we can produce similar polyhedra for the mass-deformed nonlinear sigma models on the Grassmann manifold by applying the moduli matrix formalism \cite{Isozumi:2004jc} and the pictorial representation \cite{Lee:2017kaj}. Non-Abelian junctions can be analysed by making use of the polyhedra instead of the Pl\"{u}cker embedding. We present diagrams for vacua, walls and three-pronged junctions, and compute three-pronged junction positions of the mass-deformed nonlinear sigma models on the Grassmann manifold. We show that the results are consistent with the known results of \cite{Eto:2005fm}, which are worked out by using the Pl\"{u}cker embedding. 
\end{abstract}

\end{titlepage}
\section{Introduction}  \label{sec:intro}
\setcounter{equation}{0}

The moduli matrix formalism is proposed to construct 1/2 Bogomol'nyi-Prasad-Sommerfield (BPS) walls in non-Abelian gauge
theories \cite{Isozumi:2004jc}. In the infinite coupling limit, the model becomes a massive hyper-K\"{a}hler nonlinear sigma
model on the cotangent bundle over the Grassmann manifold $T^\ast G_{N_F,N_C}$, which is defined by
$G_{N_F,N_C}=\frac{SU(N_F)}{SU(N_C)\times SU(N_F-N_C)\times U(1)}$. It is shown that the moduli space of vacua and walls is the Grassmann manifold. It is observed that walls in non-Abelian gauge theory can pass through each other.

Kink solutions of the nonlinear sigma models on $SO(2N)/U(N)$ and
$Sp(N)/U(N)$, which are quadrics in the Grassmann manifold $G_{2N,N}$ are studied in \cite{Arai:2011gg,Lee:2017kaj}. It is shown that the moduli spaces of vacua and walls are $SO(2N)/U(N)$ and $Sp(N)/U(N)$ respectively. The moduli spaces are described by pictorial representations in which the vacua and the elementary walls correspond to vertices and line segments in the representation \cite{Lee:2017kaj}. Penetrable walls appear as parallelograms in the pictorial representation and produce a recurrence of a two dimensional diagram for each $N$ mod 4 at the vacua that are connected to the maximum number of elementary walls. The structures are proved by induction.

The $1/4$ BPS states \cite{Abraham:1990nz,Oda:1999az} and domain wall webs, which contain two or more wall junctions are obtained in Abelian gauge theories and non-Abelian gauge theories by the moduli matrix formalism \cite{Eto:2005cp,Eto:2005fm}. Three-pronged junctions of the mass-deformed nonlinear sigma models on $\mathbf{C}P^{N_F-1}$, which are Abelian gauge theories are studied in \cite{Eto:2005cp}. Three-pronged junctions of the mass-deformed nonlinear sigma models on $G_{N_F,N_C}$, which are non-Abelian gauge theories for $N_C\geq2$, are studied in \cite{Eto:2005fm}. A three-pronged junction is formed by three vacua and three non-penetrable walls interpolating the vacua. To obtain the junction solutions in the Grassmann manifold with $N_C\geq 2$, which is non-Abelian gauge theory, there are two nontrivial technical complications. We should be able to identify adjacent vacua interpolated by non-penetrable walls and find the inverses of $N_C\times N_C$ matrices $S$ from non-diagonal matrices $SS^\dagger$, which are introduced in the moduli matrix formalism. In \cite{Eto:2005fm}, the Grassmann manifold $G_{N_F,N_C}$ is embedded into the complex projective space $\mathbf{C}P^{{}_{N_F}\mathrm{C}_{N_C}-1}$ by the Pl\"{u}cker embedding resolving the complications. However this method cannot be directly applied to junctions of the mass-deformed nonlinear sigma models on $SO(2N)/U(N)$ and $Sp(N)/U(N)$, which are realized as quadrics in the Grassmann manifold.

Therefore it is still useful to examine wall junctions of the mass-deformed nonlinear sigma models on the Grassmann manifold with $N_C \times N_F$ moduli matrices since the vacua and the walls are studied with $N_C \times N_F$ moduli matrices in the Grassmann manifold \cite{Isozumi:2004jc} and with $N \times 2N$ moduli matrices in $SO(2N)/U(N)$ and $Sp(N)/U(N)$ \cite{Lee:2017kaj}. 

The purpose of this work is to propose an alternative method of constructing three-pronged junctions of the mass-deformed nonlinear sigma models on the Grassmann manifold instead of using the Pl\"{u}cker embedding. We apply the pictorial representation, which is proposed in \cite{Lee:2017kaj} to vacua and walls of the mass-deformed nonlinear sigma models on the Grassmann manifold, which have been studied in \cite{Isozumi:2004jc}. We show that we can produce polyhedra, which are similar to the polyhedra \cite{Eto:2005cp} that are introduced to study BPS objects of the mass-deformed nonlinear sigma models on the complex projective space, by reformulating diagrams for vacua and elementary walls in the pictorial representation. Vertices, edges and triangular faces of the polyhedra correspond to vacua, walls and three-pronged junctions. We identify adjacent vacua that are interpolated by walls from the diagrams and construct three-pronged junctions by building polyhedra. The junction positions can be calculated from the neighbouring walls. We present some results derived from the polyhedra and show that the results are consistent with the results of \cite{Eto:2005fm}, which are obtained by using the Pl\"{u}cker embedding.

In Section \ref{sec:two}, we review the model and the moduli matrix formalism, which are studied in \cite{Eto:2005cp,Eto:2005fm}. In Section \ref{sec:three}, we present the pictorial representations of vacua and elementary walls of the mass-deformed nonlinear sigma models on $G_{N_F,N_C}$ with $(N_F,N_C)=(4,2),(5,2),(5,3),(6,2),(6,3),(6,4)$. In Section \ref{sec:four}, we study the pictorial representations for single three-pronged junctions of the mass-deformed nonlinear sigma models on $G_{N_F,N_C}$. We present diagrams for three-pronged junctions in $G_{5,2}$ as an example by reformulating the diagram for vacua and elementary walls. In Section \ref{sec:five}, we summarize our results.

\section{BPS states}  \label{sec:two}
\setcounter{equation}{0}

In this section, we review the model and the moduli matrix formalism, which are discussed in \cite{Eto:2005cp,Eto:2005fm}. The model is $3+1$ dimensional ${\mathcal{N}}=2$ supersymmetric $U(N_C)$ gauge theory with
$N_F(>N_C)$ massive hypermultiplets. $W_{\mu}$, $(\m=0,1,2,3)$ are $N_C \times N_C$ gauge field matrices, $\S_\a$,
$(\a=1,2)$ are $N_C \times N_C$ real matrices, and $M_\a$, $(\a=1,2)$ are traceless diagonal mass matrices, which are parameterized as
$M_1=\mathrm{diag}(m_1,m_2,\cdots,m_{N_F})$ and $M_2=\mathrm{diag}(n_1,n_2,\cdots,n_{N_F})$. The Fayet-Iliopoulos (FI) parameters are
$c^a=(0,0,c>0)$. We set $c=1$ in this paper. In the infinite coupling limit, the mass-deformed gauge theory reduces to the mass-deformed hyper-K\"{a}hler nonlinear sigma model on the cotangent bundle over the Grassmann manifold $T^\ast G_{N_F,N_C}$. With the FI parametrization, the fields in the hypermultiplet, which parametrize the cotangent space vanish
for the BPS equations.

The $1/4$ BPS equations \cite{Eto:2005cp} are derived by Bogomol'nyi completion of the energy density. The static configurations are constructed by setting $\p_0=\p_3=0$. We also set $W_0=W_3=0$. The BPS equations for domain wall webs on the Grassmann manifold $G_{N_F,N_C}$ \cite{Eto:2005cp} are obtained as
\bea
D_\a\ph=\ph M_\a-\S_\a \ph, \quad (\a=1,2),
\eea
where $\ph$ is an $N_C \times N_F$ matrix and $D_\m\ph=\p_\m\ph-iW_\m\ph,~(\m=1,2)$. The matrix $\ph$ is constrained by
\bea
\ph \ph^\dagger -cI_{N_C}=0, \quad (c=1).
\eea
The $1/4$ BPS system reduces to the $1/2$ BPS system when the $x^2$ dependence and the mass $M_2$ are turned off.

All the solutions of the $1/4$ BPS equations in $G_{N_F,N_C}$ can be solved by $S$ and $H_0$. $S$ are invertible $N_C\times N_C$ matrices and $H_0$ are $N_C\times N_F$ matrices. $H_0$ is called a moduli matrix.
The solution to
the BPS equation for domain wall webs \cite{Eto:2005cp} is
\bea 
\ph=S^{-1}H_0e^{M_1x^1+M_2x^2}, \label{eq:ph} 
\eea 
where 
\bea
\S_\a-iW_\a=:S^{-1}\p_\a S,\quad (\a=1,2),
\eea
and 
\bea SS^\dagger=H_0e^{2(M_1x^1+M_2x^2)}H_0^\dagger.
\label{eq:ssd} 
\eea

There is the worldvolume symmetry
\bea
H_0\to H_0^\prime=V H_0,\quad S \to S^\prime=VS,
\eea
with $V\in GL(N_C,\mathbf{C})$. Therefore the total moduli space is the Grassmann manifold
\bea
{\mathcal{M}^{\mathrm{tot}}}\simeq G_{N_F,N_C}=
\{H_0|H_0\sim VH_0, V\in GL(N_C,\mathbf{C}) \},
\eea
which includes $1/4$ BPS states, $1/2$ BPS walls and discrete SUSY vacua:
\bea
{\mathcal{M}^{\mathrm{tot}}}\simeq G_{N_F,N_C}={\mathcal{M}}_{1/4}^{\mathrm{webs}}\bigcup {\mathcal{M}}_{1/2}^{\mathrm{walls}}\bigcup {\mathcal{M}}_{1/1}^{\mathrm{vacua}}.
\eea

In Abelian gauge theories, the scalar fields \cite{Eto:2005cp} are
\bea
\ph_{A}\sim \frac{H_{0A}e^{m_Ax^1+n_Ax^2}}{\sqrt{\sum^{N_F}_{B=1}|H_{0B}|^2e^{2(m_Bx^1+n_Bx^2)}}}.
\eea
The weight of the vacuum $\la A \ra$ is defined as
\bea
\lt(H_0e^{M_1x^1+M_2x^2}\rt)_A=e^{a_A+m_Ax^1+n_Ax^2}, \label{eq:weight}
\eea
where $e^{a_A}$ is the real part of the coordinates in the moduli matrix. The position of the wall which interpolates two vacua is determined by the condition of equal weights of the vacua. The position of the wall which interpolates $\la A \ra$ and $\la B \ra$ is
\bea
(m_A-m_B)x^1+(n_A-n_B)x^2+a_A-a_B=0. \label{eq:abel_pos}
\eea

Abelian three-pronged junctions divide sets of three vacua with different labels in one color component whereas non-Abelian three-pronged junctions divide sets of three vacua with different labels in two color components. Abelian junctions exist both in Abelian gauge theories and non-Abelian gauge theories while non-Abelian junctions exist only in non-Abelian gauge theories.

In \cite{Eto:2005fm}, Abelian junctions and non-Abelian junctions of the Grassmann manifold are studied  by embedding $G_{N_F,N_C}$ into the complex projective space  $\mathbf{C}P^{{}_{N_F}\mathrm{C}_{N_C}-1}$ by the Pl\"{u}cker embedding. Therefore the wall separating $\la \cdots A\ra$ and $\la \cdots B\ra$ is on
\bea
(m_A-m_B)x^1+(n_A-n_B)x^2+a^{\la \cdots A\ra}-a^{\la \cdots B\ra}=0,\label{eq:gr_junc}
\eea
where $e^{a^{\la\cdots\ra}}$ are the real parts of the Pl\"{u}cker coordinates.

\section{Vacua and elementary walls }  \label{sec:three} 
\setcounter{equation}{0}

In this section, we apply the pictorial representation, which is used in \cite{Lee:2017kaj} to the vacua and the walls of the mass-deformed nonlinear sigma models on the Grassmann manifold, which have been discussed in \cite{Isozumi:2004jc}. Let $\la A \ra$ denote a vacuum and $\la A \leftarrow B\ra$ denote the elementary wall that interpolates $\la A \ra$ and $\la B\ra$. The moduli matrix of elementary wall $\la A \leftarrow B \ra$ in $G_{N_F,N_C}$ is $H_{0\la A \leftarrow B \ra}=H_{0\la A \ra}e^{E_i(r)}$ where $E_i(r)\equiv e^rE_i$ and $E_i$ is a simple root generator of $SU(N_F)$ \cite{Isozumi:2004jc}. Elementary walls can be identified with simple roots \cite{Sakai:2005sp}. The simple root generators and the simple roots of $SU(N)$ \cite{Isaev:2018xcg} are
\bea
&&E_i=e_{i,i+1},\nn\\
&&\vec{\a}_i=\hat{e}_i-\hat{e}_{i+1},~~(i=1,\cdots,N-1).
\eea
The matrix $e_{i,j}$ is an $N\times N$ matrix of which the $(i,j)$ component is one. The set of vectors $\{ \hat{e}_i\}$ is the orthogonal unit vectors $\hat{e}_i\cdot\hat{e}_j=\d_{ij}$.

\begin{figure}[ht!]
\begin{center}
\includegraphics[width=6cm,clip]{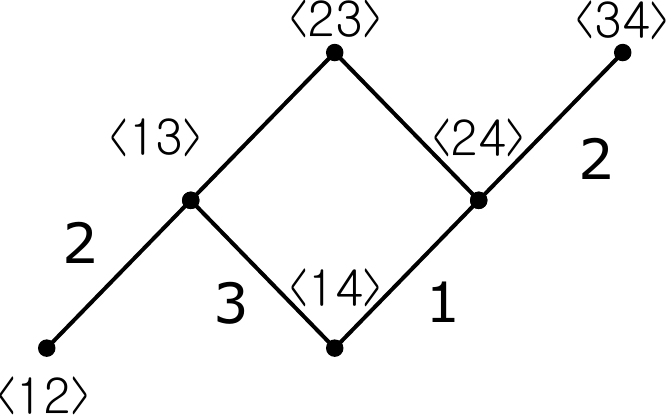}
\end{center}
 \caption{ Vacua and elementary walls in $G_{4,2}$. The numbers indicate the subscripts of the simple roots $\vec{\a}_i$.  }
 \label{fig:g42}
\end{figure}

The diagram of vacua and elementary walls in $G_{4,2}$ is depicted in Figure \ref{fig:g42}. The vertices and the line segments
correspond to the vacua and the elementary walls respectively. The parallelogram in the diagram presents two sets of
penetrable walls. A pair of facing
sides of the parallelogram are the same simple roots while a pair of adjacent sides of the parallelogram are orthogonal simple
roots.

\begin{figure}[ht!]
\begin{center}
$\begin{array}{ccc}
\includegraphics[width=6cm,clip]{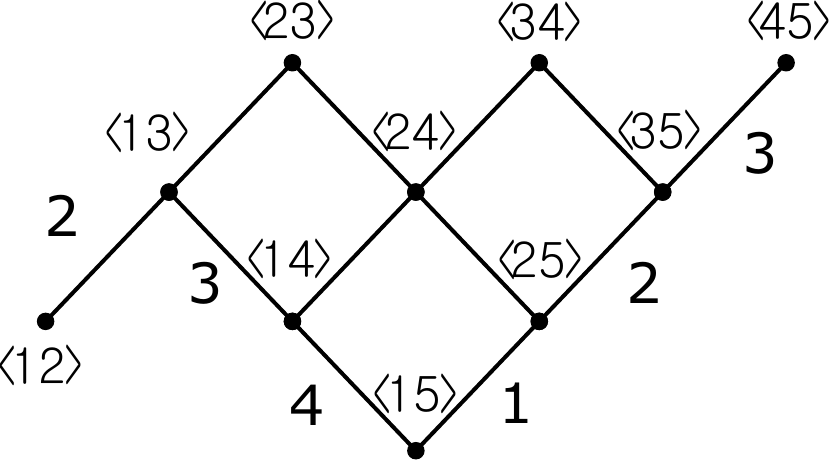}
&~~~~~~&
\includegraphics[width=6cm,clip]{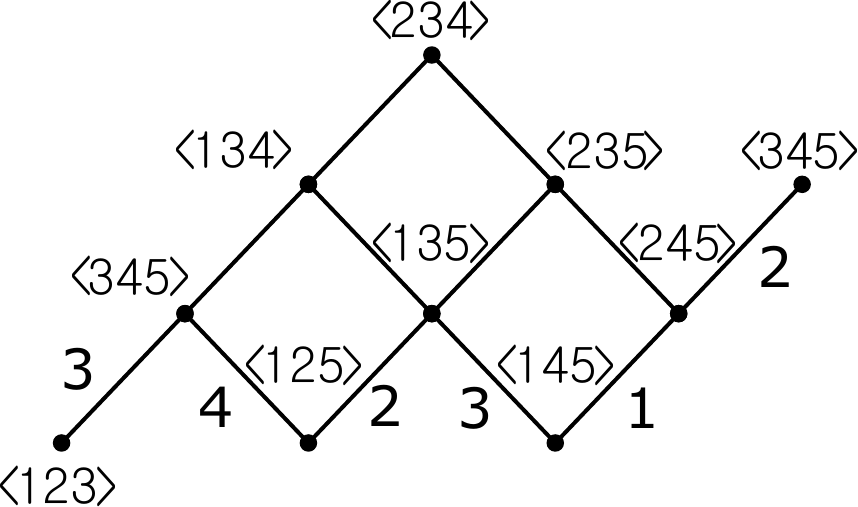}\\
\mathrm{(a)} &~~~~~~& \mathrm{(b)}
\end{array}
$
\end{center}
 \caption{ Vacua and elementary walls in $G_{5,N}$. (a)$N=2$ (b)$N=3$.  }
 \label{fig:g5}
\end{figure}

\begin{figure}[ht!]
\begin{center}
$\begin{array}{ccc}
\includegraphics[width=7cm,clip]{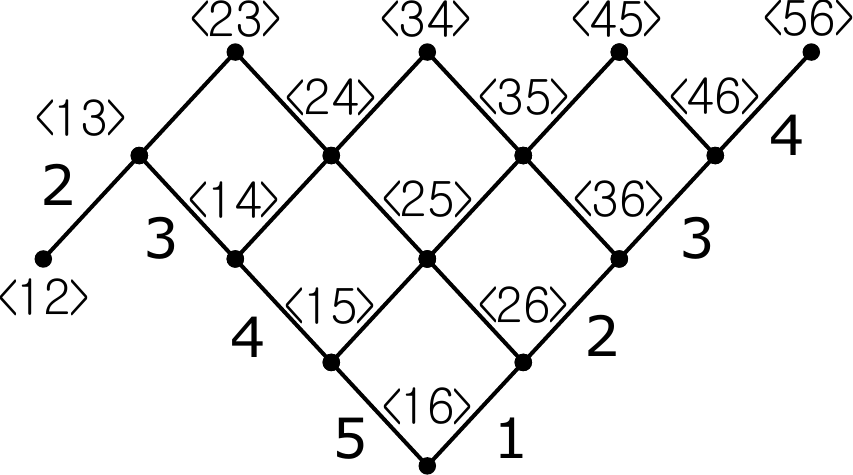}
&~~&
\includegraphics[width=7cm,clip]{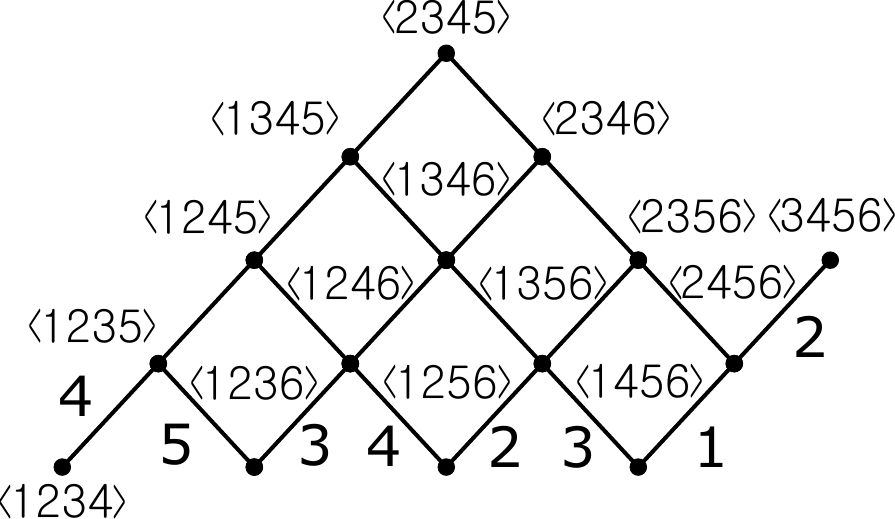}\\
\mathrm{(a)} &~~~~~~& \mathrm{(b)}
\end{array}
$
\includegraphics[width=10cm,clip]{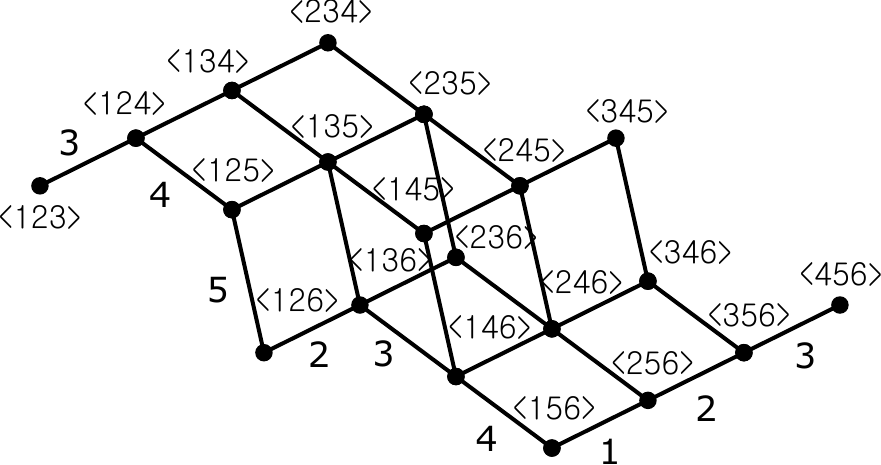}\\
(c)
\end{center}
 \caption{ Vacua and elementary walls in $G_{6,N}$. (a)$N=2$ (b)$N=4$ (c) $N=3$. }
 \label{fig:g6}
\end{figure}

The diagrams of vacua and elementary walls in $G_{5,2}$ and $G_{5,3}$ are depicted in Figure \ref{fig:g5}. Two diagrams are related by a $\pi$ rotation. This reflects the duality between $G_{N_F,N_C}$ and $G_{N_F,N_F-N_C}$. 
The diagrams of vacua and elementary walls in $G_{6,2}$, $G_{6,4}$ and $G_{6,3}$ are depicted in Figure \ref{fig:g6}. The configuration in Figure \ref{fig:g5}(a) appears in Figure \ref{fig:g6}(a).

\section{Three-pronged wall junctions}  \label{sec:four}
\setcounter{equation}{0}

We study wall junctions in the moduli space $G_{N_F,N_C}$. The moduli matrices in $G_{N_F,N_C}$ can be parameterized by real
parameters $a_{ij}$ and $b_{ij}$ as \bea
&&\lt(H_0\rt)_{ij}:=\exp(a_{ij}+ib_{ij}), \nn\\
&&(i=1,\cdots, N_C;~j=1,\cdots, N_F).
\eea
We discuss three-pronged junctions in the moduli space $G_{5,2}$. The moduli matrices in $G_{5,2}$ can be parameterized as \bea
\lt(H_{0}^{G_{5,2}}\rt)_{ij}:=\exp(a_{ij}+ib_{ij}),~(i=1,2;~j=1,2,\cdots 5), \label{eq:h052} \eea with real parameters
$a_{ij}$ and $b_{ij}$.

\begin{figure}[ht!]
\begin{center}
$\begin{array}{ccc}
\includegraphics[width=6cm,clip]{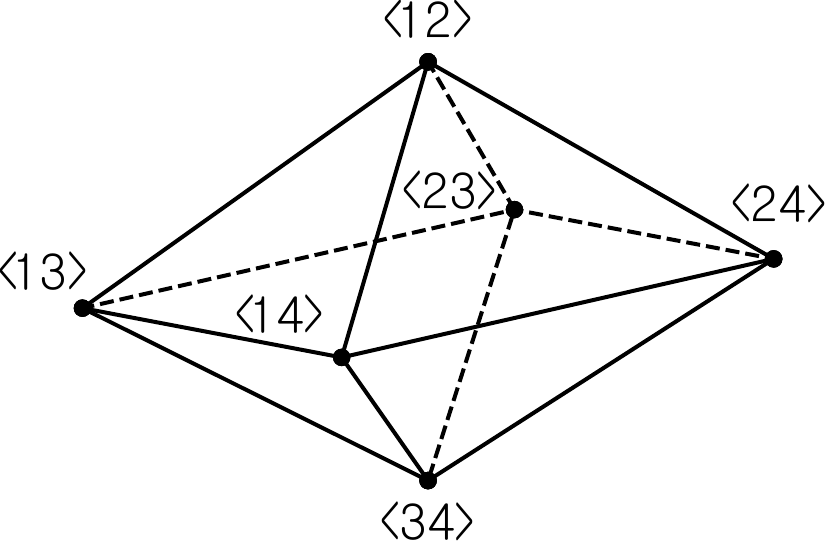}
&~~~~~~&
\includegraphics[width=6cm,clip]{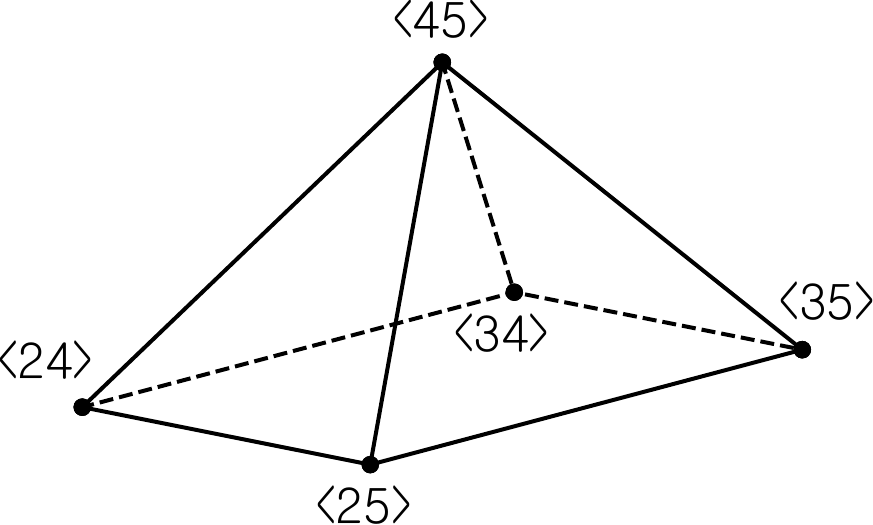}\\
\mathrm{(a)} &~~~~~~& \mathrm{(b)}
\end{array}
$
\end{center}
 \caption{Polyhedra for $G_{5,2}$  }
 \label{fig:g52_junc}
\end{figure}

Polyhedra are proposed in \cite{Eto:2005cp} to study BPS objects of the mass-deformed nonlinear sigma models on the complex projective space. Vertices, edges and triangular faces of the polyhedra correspond to vacua, walls and three-pronged junctions. We build similar polyhedra from the diagram in Figure \ref{fig:g5}(a). A single three-pronged junction is determined by three vacua which correspond to three vertices of a triangle that the wall junction gets mapped onto. We choose two sets of triangles from Figure \ref{fig:g5}(a) as shown in Figure \ref{fig:g52_junc} as an example. The diagram in Figure \ref{fig:g52_junc}(a) is an octahedron which is composed of eight triangles and the diagram in Figure \ref{fig:g52_junc}(b) is a pyramid which is composed of four triangles. The vertices, the edges and the triangular faces of the polyhedra correspond to vacua, walls and three-pronged junctions in $G_{5,2}$.

The moduli matrices for the configuration in Figure \ref{fig:g52_junc}(a) are the limits of (\ref{eq:h052}) as $a_{i5}\to -\infty$, $(i=1,2)$. There are eight triangles in Figure \ref{fig:g52_junc}(a). $\{\la 12 \ra,\la 13 \ra, \la 14 \ra\}$, $\{\la 12 \ra,\la 23 \ra, \la 24 \ra\}$, $\{\la 13 \ra,\la 23 \ra, \la 34 \ra\}$, and $\{\la 14 \ra,\la 24 \ra, \la 34 \ra\}$ are divided by Abelian junctions whereas $\{\la 12 \ra,\la 14 \ra, \la 24 \ra\}$, $\{\la 12 \ra,\la 13 \ra, \la 23 \ra\}$, $\{\la 13 \ra,\la 34 \ra, \la 14 \ra\}$, and $\{\la 23 \ra,\la 34 \ra, \la 24 \ra\}$ are divided by non-Abelian junctions. Parallelogram $\{\la 13 \ra,\la 14 \ra,\la 24 \ra,\la 23 \ra\}$ presents two sets of penetrable walls.

The moduli matrix of the Abelian junction that divides $\{\la 12 \ra,\la 13 \ra, \la 14 \ra\}$ is
\bea
H_{0\la 121314 \ra}=\lt(
\begin{array}{ccccc}
e^{a_{11}+ib_{11}} &  0           &  0             &  0   &  0 \\
0            & e^{a_{22}+ib_{22}} &  e^{a_{23}+ib_{23}}  &  e^{a_{24}+ib_{24}} &  0
\end{array}\rt). \label{eq:121314}
\eea
This is the limit of (\ref{eq:h052}) as $a_{1i}\to -\infty$, $(i=2,\cdots 5)$, and $a_{2j} \to -\infty$, $(j=1,5)$. The solution is
\bea
&&\ph=S_{\la 121314\ra}^{-1}H_{0\la 121314 \ra}e^{M_1x^1+M_2x^2} =\lt(
\begin{array}{ccccc}
\frac{f_1}{\sqrt{\D_1}} & 0 & 0 & 0  & 0 \\
0 &\frac{f_2}{\sqrt{\D_2}} & \frac{f_3}{\sqrt{\D_2}} & \frac{f_4}{\sqrt{\D_2}} & 0
\end{array}
\rt), \nn\\
&&f_1:=\exp(m_1 x^1 +n_1 x^2+a_{11}+ib_{11}), \nn\\
&&f_n:=\exp(m_n x^1 +n_n x^2+a_{2n}+ib_{2n}),\quad (n=2,3,4),
\eea
with
\bea
&&S_{\la 121314\ra}S_{\la 121314\ra}^\dagger =H_{0\la 121314 \ra} e^{2M_1x^1+2M_2x^2}H_{0\la 121314 \ra}^\dagger =\mathrm{diag}(\D_1,\D_2), \nn\\
&&\D_1:=e^{2m_1x^1+2n_1x^2+2a_{11}},  \nn\\
&&\D_2:=\sum_{n=2}^4 e^{2m_nx^1+2n_nx^2+2a_{2n}}.
\eea

The wall interpolating $\la 1A \ra$ and $\la 1B \ra$, $(A,B=2,3,4)$ is on
\bea
(m_A-m_B)x^1+(n_A-n_B)x^2+a_{2A}-a_{2B}=0.
\eea
Therefore the position of the junction that divides $\{\la 12 \ra,\la 13 \ra,\la 14 \ra\}$ is
\bea
&&(x^1,x^2)=\lt(\frac{S_1}{S_3},\frac{S_2}{S_3}\rt), \nn\\
&&~ \nn\\
&&S_1:=(-n_3+n_4)a_{22}+(-n_4+n_2)a_{23}+(-n_2+n_3)a_{24}, \nn\\
&&S_2:=(m_3-m_4)a_{22}+(m_4-m_2)a_{23}+(m_2-m_3)a_{24}, \nn\\
&&S_3:=(n_3-n_4)m_2+(n_4-n_2)m_3+(n_2-n_3)m_4. \label{eq:jp121314}
\eea

In the same manner, the moduli matrices of other Abelian junctions can be determined. The wall interpolating $\la 2A \ra$ and $\la 2B \ra$, $(A,B=1,3,4)$ is on
\bea
(m_A-m_B)x^1+(n_A-n_B)x^2+a_{1A}-a_{1B}=0,
\eea
and the wall interpolating $\la 3A \ra$ and $\la 3B \ra$, $(A,B=1,2,4)$ is on
\bea
(m_A-m_B)x^1+(n_A-n_B)x^2+a_{1A}-a_{1B}=0.
\eea

Three vacua $\{\la 12 \ra, \la 13 \ra, \la 23 \ra \}$ are divided by a non-Abelian junction. $SS^\dagger$ in (\ref{eq:ssd}) are diagonal for Abelian junctions so we can calculate junction positions by comparing weights. However, $SS^\dagger$ are not diagonal for non-Abelian junctions in general. As three-pronged wall junctions are solitons which divide three vacua interpolated by non-penetrable walls, junction positions can be calculated by finding sets of three vacua, which correspond to the vertices of the triangles that junctions are mapped onto.

We calculate the junction position from the neighbouring walls. The wall junction dividing three vacua $\{\la 12 \ra,\la 13 \ra,\la 23 \ra\}$ should be on the intersection point of the following linear equations:
\bea
&&(m_2-m_3)x^1+(n_2-n_3)x^2+a_{22}-a_{23}=0, \nn\\
&&(m_1-m_3)x^1+(n_1-n_3)x^2+a_{11}-a_{13}=0, \nn\\
&&(m_1-m_2)x^1+(n_1-n_2)x^2+a_{11}-a_{12}=0,
\eea
which are the positions of the walls  interpolating $\{\lra{12},\lra{13}\}$, $\{\lra{12},\lra{23}\}$ and $\{\lra{13},\lra{23}\}$ respectively.
The condition for the existence of the solution is
\bea
a_{12}-a_{13}=a_{22}-a_{23}.
\eea
The non-Abelian junction position is
\bea
&&(x,y)=\lt(\frac{V_1}{V_3},\frac{V_2}{V_3}\rt),\nn\\
&&~\nn\\
&&V_1:=(n_2-n_3)a_{11}+(n_3-n_1)a_{12}+(n_1-n_2)a_{13},   \nn\\
&&V_2:=(-m_2+m_3)a_{11}+(-m_3+m_1)a_{12}+(-m_1+m_2)a_{13}, \nn\\
&&V_3:=(-n_2+n_3)m_1+(-n_3+n_1)m_2+(-n_1+n_2)m_3. \label{eq:jp121323}
\eea

In \cite{Eto:2005fm}, the wall webs in $G_{4,2}$ are studied by embedding $G_{4,2}$ to $\mathbf{C}P^5$ by the Pl\"{u}cker
embedding and the junction positions are obtained from the wall webs. Since sector $\{\lra{12},\lra{13},\lra{14}\}$ and sector
$\{\lra{12},\lra{13},\lra{23}\}$ of $G_{5,2}$ reside in $G_{4,2}$ as we can see in Figure \ref{fig:g42} and Figure \ref{fig:g5}(a), we can compare the junction positions (\ref{eq:jp121314}) and
(\ref{eq:jp121323}) with the results\footnote{In \cite{Eto:2005fm}, $e^{a^{\la \cdots \ra}}$ are the real parts of the
Pl\"{u}cker coordinates. The position of the junction that divides $\{\lra{12},\lra{13},\lra{14}\}$ is
$(x,y)=$$\Big(\frac{-2a^{\la12\ra}-a^{\la13\ra}+3a^{\la14\ra}}{2\sqrt{3}},$$\frac{-a^{\la13\ra}+a^{\la14\ra}}{2}\Big)$. The
position of the junction that divides $\{\lra{12},\lra{13},\lra{23}\}$ is
$(x,y)=$$\Big(\frac{a^{\la13\ra}-a^{\la23\ra}}{2\sqrt{3}},$$\frac{2a^{\la12\ra}-a^{\la13\ra}-a^{\la23\ra}}{6}\Big)$. } of
\cite{Eto:2005fm} with mass parameters $[m_A,n_A]=\{[-\sqrt{3},-1]$,$[\sqrt{3},-1]$, $[0,2]$, $[0,0]\}$. The junction
position (\ref{eq:jp121314}) of $\{\lra{12},\lra{13},\lra{14}\}$ with these mass parameters is \bea
(x^1,x^2)=\lt(\frac{-2a_{22}-a_{23}+3a_{24}}{2\sqrt{3}},\frac{-a_{23}+a_{24}}{2}\rt), \eea and the junction position
(\ref{eq:jp121323}) of $\{\lra{12},\lra{13},\lra{23}\}$ is \bea
(x^1,x^2)=\lt(\frac{a_{11}-a_{12}}{2\sqrt{3}},\frac{a_{11}+a_{12}-2a_{13}}{6}\rt). \eea The moduli parameters in the Grassmann manifold $G_{4,2}$ are related to
the moduli parameters in the complex projective space $\mathbf{C}P^5$ of \cite{Eto:2005fm} by \bea
&&a_{22}-a_{23}=a^{\la 12 \ra}-a^{\la 13 \ra}, \nn\\
&&a_{23}-a_{24}=a^{\la 13 \ra}-a^{\la 14 \ra}, \nn\\
&&a_{11}-a_{13}=a^{\lra{12}}-a^{\lra{23}},\nn\\
&&a_{11}-a_{12}=a^{\lra{13}}-a^{\lra{23}}.
\eea
The junction positions are the same as the junction positions obtained in \cite{Eto:2005fm}.

In Figure \ref{fig:g52_junc}(b), two triangles are divided by Abelian junctions and the other two triangles are divided by
non-Abelian junctions. Parallelogram $\{\lra{24},\lra{25},\lra{35},\lra{34}\}$ is two sets of penetrable walls. The same analysis can be done on the pyramid.

We have shown that the full configurations of vacua, walls and single three-pronged junctions in $G_{N_F,N_C}$ can be determined by building polyhedra. We can always identify the corresponding $N_C\times N_F$ moduli matrices defined on the Grassmann manifold.

\section{Summary }  \label{sec:five}
\setcounter{equation}{0}

We have presented the diagrams for vacua and elementary walls of the mass-deformed nonlinear sigma models on $G_{N_F,N_C}$ with $(N_F,N_C)=(4,2)$,$(5,2)$,$(5,3)$,$(6,2)$,$(6,3)$,$(6,4)$ in the pictorial representation, which is proposed in
\cite{Lee:2017kaj}. We have observed that the duality of the Grassmann manifolds between $N_C$ and $N_F-N_C$ is realized as a $\pi$ rotational symmetry
in the representation.

In \cite{Eto:2005cp}, polyhedra are proposed to study BPS objects of the mass-deformed nonlinear sigma models on the complex projective space, which are Abelian gauge theories. Vertices, edges and triangular faces of the polyhedra correspond to vacua, walls and three-pronged junctions. In this work, we have shown that we can build similar polyhedra for the mass-deformed nonlinear sigma models on the Grassmann manifold, which are non-Abelian gauge theories, by reformulating the diagrams for vacua and elementary walls in the pictorial representation. We have analysed vacua, walls and three-pronged junctions of the mass-deformed nonlinear sigma models on the Grassmann manifold by using the polyhedra instead of the Pl\"{u}cker embedding. We have calculated junction positions of an Abelian junction and a non-Abelian junction in $G_{5,2}$ and compared the results with the results of \cite{Eto:2005fm}, which are obtained by the Pl\"{u}cker embedding. We have shown that this method produces consistent results. 

By construction, we can always identify the corresponding moduli matrices that are defined on the Grassmann manifold. Therefore the polyhedron method can be applied to the mass-deformed nonlinear sigma models on the quadrics of the Grassmann manifold such as $SO(2N)/U(N)$ and $Sp(N)/U(N)$, which are non-Abelian gauge theories for $N\geq 4$ and $N\geq 3$ respectively.

It is observed in \cite{Lee:2017kaj} that the configurations of vacua and elementary walls of the mass-deformed nonlinear sigma models on $SO(2N)/U(N)$ and $Sp(N)/U(N)$ exhibit distinguishable features as the lengths of the simple roots of $SO(2N)$ are the same whereas the lengths of the simple roots of $USp(2N)$ are not all the same. These properties would contribute to three-pronged junctions of the models since three-pronged junctions are formed by compressed walls as well as elementary walls. We hope to report on the results elsewhere.

~~\\~~\\
\noindent {\bf Acknowledgement}

This research was supported by Basic Science Research Program through the National Research Foundation of Korea (NRF-2017R1D1A1B03034222).

\end{document}